\documentclass[12pt]{article}
\usepackage{graphicx}
\usepackage{amssymb,amsmath,amsfonts,palatino,amsthm}
\usepackage{amssymb}
\usepackage{epstopdf}
\newcommand{\beq}{\begin{equation}}
\newcommand{\eeq}{\end{equation}}
\newcommand{\f}{\begin{equation}}
\newcommand{\ff}{\end{equation}}

\newcommand{\Scal}{\mathcal{S}}
\newcommand{\R}{\mathcal{R}}

\newcommand{\weave}{\left| W\rangle\right.}
\newcommand{\braweave}{\left.\langle W\right|}
\newtheorem{definition}{Definition}

\begin{document}

\title{Disordered locality in loop quantum gravity states}
\author{
Fotini Markopoulou\thanks{Email address: fmarkopoulou@perimeterinstitute.ca}\  \
and
Lee Smolin\thanks{Email address:
lsmolin@perimeterinstitute.ca}\\
\\
\\
Perimeter Institute for Theoretical Physics,\\
35 King Street North, Waterloo, Ontario N2J 2W9, Canada, and \\
Department of Physics, University of Waterloo,\\
Waterloo, Ontario N2L 3G1, Canada\\}
\date{January 15, 2007}
\maketitle
\vfill
\begin{abstract}
We show that loop quantum gravity  suffers from 
a potential problem with non-locality, coming from a mismatch between 
micro-locality, as defined by the combinatorial structures of their
microscopic states, and macro-locality, defined by the metric which emerges
from the low energy limit. As a result, the low energy limit may suffer
from a disordered locality characterized by identifications of far away points. 
We argue that if such defects in locality are rare enough they will be difficult to detect.

\end{abstract}
\vfill
\newpage


\tableofcontents

\section{Introduction}

A key issue facing loop quantum gravity is the problem of the low energy limit: how is classical spacetime and the dynamics given
by general relativity to emerge as the low energy limit of
the fundamental combinatorial dynamics. 

There are a number of approaches to this problem,
see \cite{invitation}-\cite{lowenergy}.  While there are some encouraging signs, the problem is not sufficiently undestood. One of
the most challenging obstacles to the existence of a good low energy limit  is the problem of {\it non-locality}.  

 The problem of non-locality has several aspects. One, which we focus on in this paper, arises from the fact that there are two notions of
locality in any background independent approach to quantum 
gravity in which, as in loop quantum gravity, the quantum states are
defined in terms of a combinatorial structure. There is first of all a fundamental notion of locality defined by the connectivity of the 
combinatorial structures, for example, in LQG the graphs in the spin network basis.  
We call this {\it microlocality}. 
But if the theory  has
a good classical limit, then there will be an emergent classical spacetime geometry,
defined by a metric $q_{ab}$.  Distance measurements in this metric define 
a second notion of locality, which we will call 
{\it macrolocality}.    In some discussions of the low energy limit it is assumed that the two notions of
locality will coincide.  This is done for example, by associating to a classical metric $q_{ab}$ semiclassical states with support on graphs
which are embedded in space in such a way that only nodes that are within a few Planck distances of each other, as measured by
$q_{ab}$, are connected \cite{weaves1,weaves-luca}.
The most important point in this paper is that there is no reason to make this assumption.  As we show in the next section,  the conditions commonly imposed on semiclassical states can be satisfied without the assumption that micro and macro locality coincide.   This implies that there may be non-local effects present to some extent in these definitions of  the low energy limit of loop quantum gravity. We call this phenomenon
{\em disordered locality.}

 To understand the effects of non-locality  we have studied 
 how the problem arises in detail in loop quantum gravity.  One can, for example,
 investigate states which are
constructed as follows: we begin with well-studied  semiclassical states, which are defined so that they coincide in a coarse grained sense with a given classical
metric $q_{ab}$.  We then 
contaminate the state by adding a small proportion of links that are non-local in the
classical metric $q_{ab}$ without, however, weakening the correspondence between
expectation values of coarse grained observables and the classical metric.  (See
Figure 1.)
As we will show in the next section, the fact that we can do this in itself is
evidence that the two notions of locality need not coincide.  

\begin{figure}
\begin{center}
\includegraphics[height=2.5cm]{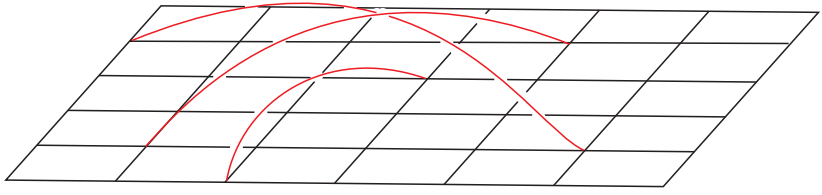}
\end{center}
\caption{A lattice contaminated by non-local links.}
\end{figure}

Our investigation of the role of non-locality in LQG will be based on two notions of non-local links. The first of these is combinatoric, the second depends on the volume observables. In most of our examples both will be satisfied.  

\begin{enumerate}

\item{} If the smallest closed cycle of links that a link  $e$ is a part of has a number of links $>>> 1$ we say that $e$ is a {\it combinatorially non-local link} \cite{hal}.

\item{} Suppose that two nodes $p$ and $q$  in a spin network $\Gamma $ have the property that every generic region 
$\cal R$ that includes both of them for which 
$\langle \Gamma | \hat{V}[{\cal R} ] | {\Gamma } \rangle $, the expectation value of the volume operator for the region $\cal R$,  is non-zero, that quantity is actually very large in Planck units.  Then we can say that $p$ and $q$ are {\it geometrically far separated.}  If there is an edge
$e$ joining $p$ and $q$ then that edge is {\it metrically non-local}.  
(By generic we mean that small deformations of $\cal R$ do not result in large changes in $\langle \Gamma | \hat{V}[{\cal R }] |{ \Gamma }\rangle$.) 

\end{enumerate}
 
In the next section we study show that disordered locality is a generic feature of states in loop quantum gravity. In 2.1 we study kinematical states associated with classical three metrics, called weave states, and show that locality may be disordered while still satisfying the conditions of correspondence to a classical metric.  In 2.2 we show that disordered locality is also generic in physical states constructed according to Thiemann's procedure given in \cite{thomas1}.

In section 3 we discuss implication of these results.  We discuss whether dynamics could suppress non-locality and enforce a match of macroscopic and microscopic locality and find evidence to the contrary.  We then discuss in 3.3 whether any density of non-local links 
kills the theory. We argue that so long as the non-local links are
sufficiently uncommon, explicit non-local interactions will be 
difficult for local observers to detect as local effects.

Before going onto the body of the paper we note that the phenomena
of disordered locality is not necessarily special to loop quantum gravity. It may arise in other background independent approaches to quantum gravity based on combinatorial structures, such as causal sets and
causal dynamical triangulation models.  But in this paper we restrict
our considerations to loop quantum gravity.

\section{The low energy limit and the problem of non-locality}

Loop quantum gravity gives an exact microscopic description of spatial quantum geometry in terms of basis states called spin networks, which are diffeomorphism classes
of certain labeled graphs.  The dynamics is expressed, in both path integral and hamiltonian formulations, in terms of amplitudes for local moves in the graphs.
Hence the dynamics of the theory respects a notion of microlocality.  

There are several proposals for the construction of candidate semiclassical states in loop quantum gravity.  The earliest approach, called weaves\cite{weaves1,weaves-luca}, requires that expectation values of coarse grained geometric observables such as areas and volumes of large regions, agree with those computed from some classical background metric $q_{ab}$. 
Other constructions are based on coherent states\cite{coherent}
 or on the exponentiation of Hamilton-Jacobi functions from classical general relativity\cite{semi-hj}.  In these cases there are results which suggest that  when coupling to the usual kinds of matter fields are included, the low energy behavior could be  described by an effective quantum field theory for the matter fields on the classical metric that describes the averaged or coarse grained 
geometry\cite{effectivematter,lqg,invitation,semi-hj}.  

There are at least two ways that the fundamental microscopic locality and emergent macroscopic
locality can differ.  First, in a single basis spin network state, an averaged coarse grained
notion of distance, that would give rise to an emergent classical metric $q_{ab}$, can
be insensitive to a small number of non-local links connecting nodes far away in the coarse
grained notion of distance.  

The second problem arises from the fact that macroscopic quantum states are likely to
involve superpositions of the spin network states. 
Indeed, as the dynamics modifies the graphs by local rules, any ground state or, more generally, time independent or physical state (a solution of all the constraints), will necessarily  involve superpositions of the spin network basis states. As such it is to be expected that macro-locality, as emergent from these states, will reflect  a quantum  averaging of micro-locality in each state in the superposition.  But if the ground state involves a superposition of many spin-network states, it is reasonable to question whether the micro-locality of each member of the superposition 
making up the semiclassical state  will agree  with the macro-locality of the resulting superposition.  If this is the case there may be correlations and processes which are local in the microscopic
structure of the Planck scale theory, but non-local in the metric which emerges in the low energy limit\cite{nfs}.

\subsection{Non-locality in weave states}

In this section we discuss the first issue of non-locality, which is a disordering of 
non-locality in a given spin network basis state.  
We will illustrate this  issue  in the simplest case in which it occurs, which are the  weave states \cite{weaves1}.  
A weave is a kinematical state of loop quantum gravity, designed to match a 
given slowly varying
classical spatial metric $q_{ab}$.   For simplicity we discuss weaves
for the spatial manifold $\Sigma=T^3$, the three torus, with standard flat metric, 
of volume $R^3$, with $R \gg l_{{\rm Pl}}$.  
In loop quantum gravity there are volume operators $\widehat{V}[\R]$ for every
region $\R$ of $\Sigma$, and area operators $\widehat{A}[\Scal]$ for every surface
$\Scal$.  For each region and surface, let $v[\R]$ and $a[\Scal]$ denote
their volumes and areas with respect to the flat metric on $T^3$.  We consider
surfaces and regions whose classical areas and volumes are large, 
and also such that the curvatures of surfaces and boundaries is small, both in 
Planck units.  A weave state for the flat metric, $\weave$ is usually defined to be one that
satisfies, for all such regions and surfaces \cite{weaves1}, 
\beq
\begin{array}{rl}
\widehat{A}[\Scal] \weave  =&  \left ( a[\Scal] + O( l_{\rm {Pl}}^2 )  \right ) \weave
 \\
\widehat{V}[\R] \weave =&  \left ( v[\R] + O( l_{\rm{Pl}}^3 )  \right ) \weave.
\end{array}
\label{weave} 
\eeq
We require here that the surfaces are generic, {\em in that they can be freely varied without changing the
area eigenvalues very much.}  This eliminates cases of measure zero where a surface goes through many spin network nodes.  

It is of course, not going to be the case that the ground state is an eigenstate of operators that measure the three geometry. 
As an eigenstate of the Hamiltonian constraint-or Hamiltonian for a fixed gauge, the ground state is going to be a superposition 
of three-geometry eigenstates, with minimal uncertainty in three geometry and extrinsic curvature variables.  
We can then weaken the weave condition (\ref{weave}) to a condition on a general candidate for a ground
state $ \weave$,  which relates the expectation value of geometric observables to the classical geometry. With the
same conditions as to large areas and volumes we require
\beq
\begin{array}{rl}
\braweave \widehat{A}[\Scal] \weave =&  a[\Scal] + O\left( l_{\rm{Pl}}^2 \right) 
 \\
\braweave\widehat{V}[\R] \weave=&    v[\R] + O\left(  l_{\rm{Pl}}^3  \right).   
\end{array}
\label{weave3}  
\eeq

This  volume condition can be satisfied by building the weave state on a regular lattice, 
or on a triangulation of space, where the lattice spacing is on the order of the Planck length. 
To satisfy the area condition for all slowly varying surfaces requires a random weave,
as constructed by \cite{weaves-luca} (See Figure 2).

\begin{figure}
\begin{center}
\includegraphics[height=3cm]{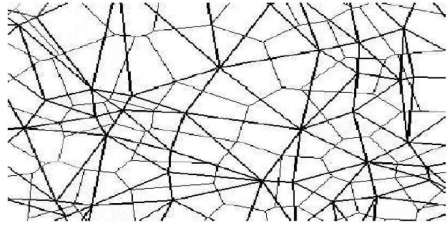}
\end{center}
\caption{A random weave from \cite{weaves-luca}.}
\end{figure}

Most weave states in the literature, including the  states constructed in \cite{weaves-luca}  satisfy an additional, unstated, condition of locality:

\begin{definition}
{\rm A weave state for a slowly varying metric $q_{ab}$ 
is a {\em local weave state} if all of the edges in the spinnetworks with non-zero
amplitude in the state connect two nodes whose metric distance measured with $q_{ab}$ is
on the order of the Planck length.}
\end{definition}

It has been shown that if one expands a theory with matter around such a local weave state,
one can derive a version of the matter quantum field theory on the background $q_{ab}$, but
with a Planck scale cutoff \cite{effectivematter}.  

Simple examples of states satisfying (\ref{weave}) have been  constructed in 
\cite{weaves-luca}.   For the simplest case of flat space one can  take a 
simplicial decomposition, $\cal T$, 
 of a large  region of $R^3$  in terms of tetrahedra, based on a random
selection of nodes, chosen with respect to the volume element formed
from $q_{ab}$.  
The dual spin network \cite{fotini-causal} is a 4-valent 
graph $\Gamma$.

For our purposes, it will be sufficient to  first consider simply a  regular weave, 
based on a regular lattice, as shown in Figure 3.  We show that the volume measurements
and area measurements in the planes of the regular lattice are unchanged if the local weave
is replaced by a non-local weave. 

We can choose to label each edge by the same spin, $j$, and each vertex by the same intertwiner, $I \in {\cal V}_{jjjj}$,
where $ {\cal V}_{jjjj}$  is the linear space of maps $I: j \otimes j \otimes j \otimes \otimes j \rightarrow id$.  The
corresponding basis state is $|\Gamma, j, I \rangle$.  To satisfy the weave conditions
we can construct a  superposition of such states based on a single lattice, 
\f
\weave = \sum_{j=\frac{1}{2}}^{\infty}\quad \sum_{I \in {\cal V}_{jjjj} } a_{j,I}  | \Gamma, j, I \rangle,
\ff
where
\f
\sum_{j=\frac{1}{2}}^{\infty}\quad \sum_{I \in {\cal V}_{jjjj} } |a_{j,I} |^2 =1.
\ff
It is straightforward to apply Bombelli's construction in \cite{weaves-luca} to show that the condition
 \ref{weave3} is satisfied if
\f
\frac{\braweave \widehat{\sqrt{j(j+1)}}  \weave }{\braweave \widehat{v} \weave} = \pi^{4/3}\left (  
\frac{7^4 2^{20}}{3^5 5^2}
\right )^{1/9}=: C.
\label{c-condition}
\ff
where $\widehat{v}$ is the combinatorial part of the volume eigenvalue.

A simple class of spin nets that satisfy this is made from only  $j=1$ edges, with a superposition of $ I=0$ and $I=1$
nodes\footnote{In passing it is interesting to note that \ref{c-condition} cannot be satisfied for low spins for a weave volume 
eigenstate.}.  Here we use the fact that the volume eigenvalues for the $j=1$ case are
$v= (0,\frac{3^{1/4}}{4}, \frac{3^{1/4}}{2})$ corresponding to $I=(0,1,2)$.  The condition \ref{c-condition} yields
\f
 \frac{|a_{1,0}|^2}{|a_{1,1}|^2} = \frac{C}{4 3^{1/4}}-1.
\ff

We note that this state is a local state in the sense just defined.  

\begin{figure}
\begin{center}
\includegraphics[height=2.5cm]{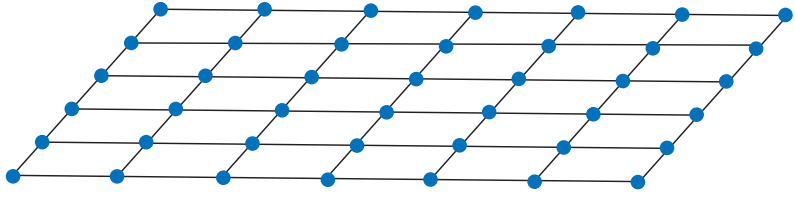}
\end{center}
\caption{A regular local weave $|\Gamma, j=1, I=1\rangle$.}
\end{figure}

\begin{figure}
\begin{center}
\includegraphics[height=2.5cm]{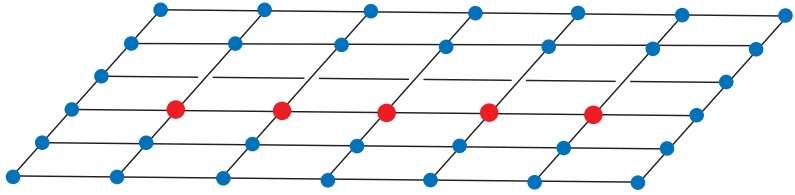}
\end{center}
\caption{A regular weave with non-local links, $|\Gamma',j=1, I=1\rangle$.}
\end{figure}

However, it is easy to see  that 
the  condition, (\ref{weave3})  of correspondence with a classical geometry is not strong enough to guarantee locality.  We can make a simple modification in the state just discussed
that preserves the weave conditions  (\ref{weave3}) for a flat 
metrics while violating the condition that micro and macro locality coincide.   This is
illustrated in Figure 4, in which we begin with 
the state $\weave$ just constructed and modify the $|\Gamma, 1,1 \rangle$ term in 
the superposition as follows.
Pick a connected chain $\alpha $ of nodes and edges of $\Gamma$ beginning 
on a node $A$ and 
ending on a node $B$ such that the Euclidean distance between them is very large in Planck units.  
Replace every node in the chain by a simple crossing as in Figure 4.  For each node eliminated, pick an
adjacent node and raise its intertwiner from the $1$ state to the $2$ state.  The result is the state
we will call $|\Gamma', 1,1\rangle$. Now replace $|\Gamma, 1,1 \rangle$ by $|\Gamma', 1,1\rangle$ in the
definition of $\weave$, leaving the amplitudes fixed. The conditions ((\ref{weave3})) are still satisfied. The areas of
all generic surfaces are unchanged. The volumes of all large regions are also unchanged, for all we have
done is eliminated a set of nodes each carrying volume $\frac{3^{1/4}}{4}$ and put their volume on adjacent
nodes by doubling their contributions to $\frac{3^{1/4}}{2}$.   We may call such a state, which satisfies (\ref{weave3})
but fails the condition of locality, a {\it non-local weave state.}

It is easy to see that there are many more non-local than local weave states of this type.  To see this note that, given any  local weave
state,  the construction just described can be carried out for a large 
number of  chains $\alpha$ in $\Gamma$, each of arbitrary length. So long as only
a fraction of the nodes are affected, the weave condition (\ref{weave3}) will still be satisfied. 
This shows that the condition of correspondence with a classical metric, (\ref{weave3}) is too
weak to guarantee locality. To put this another way, locality is a rare and unstable property of
such weave states. 

The same argument can be applied to random local weave states such as the ones studied
in \cite{weaves-luca}.

The problem of locality concerns the low energy limit because the 
 same methodology that yields local quantum field theory when expanding around a 
local weave gives a matter theory with non-locality when expanded around a non-local weave state.  
So it cannot be claimed that the condition (\ref{weave3}) by itself leads to a recovery of local quantum
field theory in the low energy limit.  

\subsection{Non-locality in solutions to all the constraints of loop quantum gravity}

One response to the problems just raised is that they will be resolved by the
dynamics. Since the weave states just studied are not solutions to the constraints
of quantum gravity one might hope that the problem does not arise for solutions
to the Hamiltonian constraint.  It is easy to see that this hope is not fulfilled.  

Thiemann, in \cite{thomas1} showed how to construct infinite numbers of
solutions to the Hamiltonian constraint. Given a root graph, he
describes an iterative procedure that constructs from it a superposition of spin network states that solves all the constraints of quantum general relativity.

But the original root graph can be any basis state in the theory. In particular, it
could be the local weave states shown in Figure 3.   But it could also 
be the non-local weave state in Figure 4. In fact, using the procedure 
followed there one can show the following: {\it Suppose we are given a  root state, that when acted on by Thiemann's procedure produces a solution to all the constraints that is also semiclassical (on any criteria desired) in which the
microscopic and macroscopic notions of locality coincide.  Then we can modify
the root state by the addition of a non-local link, and by acting produce a
new state, every member of the superposition of which has a non-local
coupling, in the macroscopic notion of locality.}

One might ask if the mismatch of microscopic and macroscopic locality could somehow wash out in the superposition involved in Thiemann's procedure?  it is easy to see that this will not usually be the case.  

Suppose $|{\cal T}\rangle$ is a Thiemann state constructed from a root graph $| r \rangle$.   Suppose also that $|{\cal T}\rangle$ satisfies  
(\ref{weave3}) for a classical slowly varying metric $q^0_{ab}$. 

Consider two nodes $p$ and $q$ of $|r \rangle$ which are geometrically far separated, according to the definition in the introduction.  They persist as nodes in every graph in the superposition constructed by Thiemann's procedure.  Furthermore, in all  of the states associated with the graphs, $p$ and $q$ will also be geometrically far separated, because the procedure may alter the expectation value of the volume at nodes that were in the root graph, but only by factors of order unity, while the number of nodes that contributes to the superposition does not change.   

 We now modify 
$|r \rangle $ by adding a link between $p$ and $q$ giving us 
$|r^\prime \rangle$ and construct from it the  Thiemann state 
$|{\cal T}^\prime \rangle$.  This will change expectations of volumes and areas by at most a few Planck volumes, so  
$|{\cal T}^\prime \rangle$ still satisfies 
\ref{weave3} with the same $q_{ab}^0$.   As a result, 
$p$ and $q$ still satisfy in the condition of being geometrically far separated in  every basis state in the superposition that constructs
$|{\cal T}^\prime \rangle $.  But they are connected by a link in every state in the superposition
$|{\cal T}^\prime \rangle$.  Hence micro-locality and macro-locality do not agree and locality has been  disordered.  

Given that we can do this to any two nodes in a root graph of a Thiemann state we see that disordering of locality is generic in physical states constructed according to Thiemann's procedure.

\section{Implications of disordered locality}

We have shown that states which are semiclassical but nonetheless have disordered locality are common in the kinematical Hilbert space of loop quantum gravity as well as among solutions to the quantum constraints.  

We should emphasize that given our results, it is the case that almost every state that satisfies the conditions (\ref{weave3}) has disordered locality.   Consider a local weave state, $|L\rangle $ with 
$N \approx \frac{\hat{V}[\Sigma ]}{l_{Pl}^3}$ nodes. Consider the effect of a small perturbation of the state gotten by adding a new edge connecting two of those nodes, using the construction described above that does not change the expectations of volumes.  There are roughly $N$ ways to do this that leave it a local weave state.  There are a much larger number
of the order of $N^2$ changes that introduce a disordering of locality, because the two nodes are far away in the classical metric.   Thus, to choose a weave state out of all the states that satisfy 
(\ref{weave3}), that is in addition local is an extremely unnatural requirement, it amounts to the suppression of on the order of
$N^2$ degrees of freedom. 

Note that while we have found disordered locality around superpositions of states produced by Thiemann's construction, we have not here addressed the general problem of non-locality in superpositions of states not constructed on the same roots.  This is addresssed 
elsewhere\cite{nfs}.

In this section,  we mention several implications of non-locality which are worthy of further investigation.  We start with dynamical questions.  We consider both the dynamics of loop quantum gravity given by local moves in which nodes are replaced by triangles and vice versa and the dynamics of spin foam models which involve a more general set of moves dual to Pachner moves on triangulations. 

\subsection{Could dynamics produce non-local links?}

Suppose we start with a weave state with no disordered locality, so that links are only connected in the graph when they are Planck separated in the classical metric.  Then under the dynamics considered in loop quantum gravity, which consists of the replacement of nodes by triangles, or the reverse \cite{thomas1}, one can not by a small number of local moves create non-local links. Nor can this be achieved by a small number of exchange moves involving two or three connected nodes as in the dual Pachner moves that characterize spin foam models\cite{lqg}. 

However, there is nothing to prevent non-local links from being created by a large number of evolution moves which include the exchange moves.  Consider two nodes $p$ and $q$ of a local wave state $\Gamma^0$,  corresponding to a classical metric $q_{ab}$ connected by an edge $e$.  There is no local move that eliminates $e$ without eliminating or merging both $p$ and $q$.  So let us consider an exchange move which leaves $p$ alone but involves $q$ being exchanged with another node $r$.  After that move, $q$ and $r$ are eliminated and two or three new nodes are created, which we can call $s_i$.  The edge 
$e$ now connects from $p$ to one of the new nodes, lets call it $s_1$.  

We can then repeat and now consider an exchange move involving $s_1$ and not $p$.  Again a new node is created, that is linked to $p$ by $e$.  Lets call it $s^2_1$ to denote that it is the first node created by the second exchange move. After a long series of $N$ such moves, not involving $p$, the edge $e$ now runs from $p$ to a node created by the last exchange move
$s^N_1$.  

If this sequence of moves is all that has ocurred then the resulting state, with $e$ excluded, may still satisfy \ref{weave3} with respect to $q_{ab}$.  But after $N$ steps $e$ could connect two nodes that are order $l_{Pl}N$ apart as measured by $q_{ab}$.  Even if the exchange moves were randomly chosen just to never involve $p$ but always to involve the other endpoint of $e$ we would expect that that end point makes a Brownian motion and is so of order $\sqrt{N} l_{Pl}$ from $p$ as measured by $q_{ab}$.   Thus, a long series of local moves can introduce a non-local connection to an otherwise local weave state. 
This was also found to occur in numerical simulations of stochastic graph evolution\cite{hal}. 

\subsection{Could dynamics eliminate non-local links?}

Even given the results just presented, it  might nevertheless be the case that dynamics would pick out states with a small or vanishing mismatch of micro- and macro-locality.   If we had under control
a hamiltonian operator, in the presence of appropriate boundary conditions or gauge fixing, which had been shown to be positive
definite, we would be able to find the ground state and compute whether locality is disordered or not.  

In the absence of such a hamiltonian, one can at least try to see if the kind of local evolution rules we find in LQG have the tendency to eliminate non-local links.  
While it is difficult to study quantum evolution numerically, one thing that can be done is to study the effects of these local rules when used to generate a classical statistical ensemble of graphs.  This was studied by Finkel \cite{hal} who performed numerical experiments in which he started with random graphs with around $200$ nodes and grew them by two orders of magnitude by applying the kinds of local rules used in loop quantum gravity randomly.   Non-locality was measured using the combinatorial definition of non-locality given first in the introduction.   This has the advantage of not needing to make reference to a classical metric.  Under a range of initial conditions and weights for different kinds of moves, Finkel saw no evidence that non-local links were eliminated by the dynamics.  Instead he found evidence that the graphs approach states in which there are fixed proportions of non-local edges.  This can also be underestood heuristically, as discussed in \cite{darkenergy}.  

These simulations also showed that such non-local links can be created by a long series of node to triangle moves\cite{hal}.  Finally, once non-local links exist, there are exchange moves involving them which will produce more non-local links, as shown in \cite{hal,darkenergy}.  

These results imply that locality is stable in the short term in that non-local links will not suddenly spring up by local moves, but locality can be disordered over times scales large compared to the Planck time. Moreover, there is no evidence that the dynamics in LQG suppresses or eliminates non-local links, once they are present.  

\subsection{Detecting non-locality}

How easy would it be to detect a small amount of disordered locality?   

Let us consider a regular local weave disordered by a small number of randomly placed nonlocal links, as in
Figure 1.  Let  $P$ be the probability that two nodes of a graph with significant weight in a semiclassical state which are far from each
other in the classical metric $q_{ab}$ are connected. Note that there will be roughly $10^{180}$ nodes within the present
Hubble scale.  This means that there are roughly $10^{360}$ potential non-local
links.  Suppose that only $10^{80}$ of these are turned on.  This means that 
there are as many non-local links as  baryons in the universe.  The 
probability $P$ is still as low as $10^{-280}$.   Even if there is one node with a 
non-local link in every cubic fermi, we are still at a very low $P \approx 10^{-120}$.   The point is that $P$ can be very small, but the 
density of nodes which are ends of non-local links can be much
higher than the density of baryons or photons.  

Suppose these non-local links are there.  How would we detect them?  This of course
depends on the dynamics. If each end of a non-local link has a Planck mass, then they will
dominate the universe. But suppose there is no energy cost, is there any way of
detecting the resulting non-locality directly?

How likely is it to have non-local effects that could be measured in
 a laboratory?
If the non-local links are randomly distributed, then the probability of having
two nodes within one laboratory connected by a non-local link is extremely small.
Let us assume that we could detect non-locality in a laboratory if two nodes within
a kilometer are connected.  But with $P$ chosen as above, the probability of there being a non-local link contained
within a kilometer cubed of volume is $10^{-44}$.  Almost every non-local link originating
within a terrestrial laboratory connects a node in the laboratory to a node somewhere in the universe outside our galaxy.  

Thus, it is going to be very unlikely with these numbers that non-locality can be directly
detected.  The effect of the non-locality will then be to introduce a randomness, which
disorders local physics by connecting regions of the universe randomly and very
weakly with distant regions.  Since the universe is already roughly in thermal equilibrium,
it would not be easy to detect the consequences.

We can estimate  the rate that a photon with energy $E$ will jump across a non-local link within the support of its wavefunction on dimensional grounds to be $GE^3$, where $G$ is Newton's constant.  Given that almost all the photons
in the universe are already in thermal equilibrium, it is not easy to see how adding a small rate for such non-local jumps to take place will be detectible even up to densities such as one mouth of a non-local link per cubic fermi.  It would be easier to detect the jumping of charged particles, however it is not clear that they can jump as charge conservation requires that the charge remains at the mouth of the wormhole the particle jumps through.  

Another way to study the consequences of disordered locality is to consider the effects on a lattice quantum
field theory of disordering it by the addition of a very small number of non-local links.  Such models are called small world networks, and
have been studied by statistical physicists.  
A study of the effect on the two-dimensional Ising model of 
the addtion of a small number of randomly placed non-local links found only a slight rise in the Curie temperature by an amount proportional
to $P$ \cite{eaton}.  At  the same time, the correlation functions
seem to scale with the temperature in such a way that, for small $P$,  non-locality is  undetectable by measurements of the correlation
functions.  

Given that there is a range of $P$ such that disordered
locality would not be easily detectible, we can go on to investigate
whether there might be consequences for cosmological problems such as dark energy or super horizon correlations.  This will be the subject
of future papers\cite{darkenergy}.

\section*{Acknowledgements}

We would like to thank Luca Bombelli for helpful comments on a draft of this paper. 
We are grateful also to Lucien Hardy, Jacob Foster, Hal Finkel, 
Joao Magueijo,  Isabeau Premont-Schwarz, Yidun Wan and Antony Valentini for conversations  and encouragement.  Research at
PI is supported in part by the Government of Canada through NSERC and
by the Province of Ontario through MEDT.

\end{document}